\documentclass[a4paper,10pt]{article}
\usepackage[a4paper,top=2.5cm,bottom=2.5cm,left=2.5cm,right=2.5cm,marginparwidth=1.75cm]{geometry}
\usepackage[pdftex, pdftitle={Article}, pdfauthor={Author}]{hyperref}
\usepackage[utf8]{inputenc}
\usepackage{multicol}
\usepackage{float,array,multirow}
\usepackage{subcaption}
\usepackage{todonotes}
\usepackage{amsmath}
\usepackage{amssymb}
\usepackage[english]{babel}
\usepackage{xcolor}
\usepackage[percent]{overpic}
\usepackage{helvet}   % in preamble
\usepackage{pifont}

\newcommand{\diiid}{\mbox{DIII-D}}
\usepackage[backend=biber,maxnames=2,style=numeric-comp,sorting=none]{biblatex}
\DeclareFieldFormat[article]{pages}{#1}
\DeclareBibliographyAlias{article}{std}
\DeclareBibliographyAlias{online}{std}
\DeclareBibliographyAlias{book}{std}
\DeclareBibliographyDriver{std}{
  \usebibmacro{bibindex}
  \usebibmacro{begentry}
  \usebibmacro{author/editor+others/translator+others}
  \setunit{\labelnamepunct}\newblock
  \newunit\newblock
  \usebibmacro{date}
  \newunit\newblock
  \usebibmacro{journal}
  \newunit\newblock
  \printfield{volume}
  \newunit\newblock
  \printfield{pages}
  \newunit\newblock
  \usebibmacro{finentry}}

\usepackage{lscape}
\usepackage{rotfloat}

\usepackage{hyperref}

\usepackage[section]{placeins}

\usepackage{graphicx}

\newlength{\figH}\newlength{\cutA}\newlength{\cutB}

\begin{document}
\clearpage\thispagestyle{empty}
\begin{center}
%\section*{Momentum transport characteristics across\\turbulence regimes in the DIII-D tokamak}
\section*{Dependence of Momentum Transport on the\\Dominant Turbulence Regime in the DIII-D Tokamak}

C. F. B. Zimmermann\textsuperscript{a b *}, C. Chrystal\textsuperscript{c}, E. Perez\textsuperscript{d}, T. Tala\textsuperscript{e}, C. Angioni\textsuperscript{b}, S. Haskey\textsuperscript{f},\\F. Khabanov\textsuperscript{d}, R. M. McDermott\textsuperscript{b}, G. McKee\textsuperscript{d}, A. Salmi\textsuperscript{e}, L. Schmitz\textsuperscript{g}\\
\vspace{0.4cm}
\textit{\textsuperscript{a} Columbia University, New York, USA}\\
\textit{\textsuperscript{b} Max Planck Institute for Plasma Physics, 85748 Garching, Germany}\\
\textit{\textsuperscript{c} General Atomics, San Diego, USA}\\
\textit{\textsuperscript{d} University of Wisconsin, Madison, USA}\\
\textit{\textsuperscript{e} VTT, P.O. Box 1000, FI-02044 VTT, Finland}\\
\textit{\textsuperscript{f} Princeton Plasma Physics Laboratory, Princeton, USA}\\
\textit{\textsuperscript{g} University of California Los Angeles, Los Angeles, USA}\\
\vspace{0.4cm}
\textsuperscript{*}\,E-mail of the corresponding author: benedikt.zimmermann@columbia.edu
\end{center}
\begin{abstract}
\noindent Accurate prediction of toroidal plasma rotation is essential for the optimization of confinement and stability in future fusion devices, where external torque sources will be weak. This work investigates turbulent core momentum transport in the DIII-D tokamak across a transition from ion-temperature-gradient (ITG)-dominated to trapped-electron-mode (TEM)-dominated turbulence. A momentum transport analysis framework previously developed for ASDEX Upgrade is applied to modulated neutral beam injection experiments, enabling the separation of diffusive, convective, and residual-stress contributions through Fourier analysis of the plasma rotation response. The studied dataset spans low-rotation conditions, dominant electron heating, and background $E\times B$ shearing rates lower than turbulence growth rates, thereby accessing more reactor-relevant conditions. Gyrokinetic CGYRO and gyrofluid TGLF calculations indicate that the parameter scan indeed covers a transition from ITG-dominated to strongly TEM-dominated turbulence. The analysis yields Prandtl numbers that vary around unity. The pinch number shows no explicit dependence on the ITG-TEM transition and is instead roughly ordered by the logarithmic density gradient. In contrast, the normalized residual stress exhibits a non-monotonic, V-shaped dependence across the ITG-TEM transition, with co-current values in deep ITG and deep TEM regimes and near-zero or counter-current values in the intermediate mixed-mode regime. This trend collapses onto an approximately linear dependence when plotted against the electron kinetic profile gradients, suggesting the presence of residual stress generation by profile-shearing effects. In addition, weaker background $E\times B$ shearing shifts the residual stress toward counter-current values. Linear CGYRO simulations for the representative ITG and TEM discharges yield Prandtl and pinch numbers in good agreement with the experimental analysis, supporting the applicability of gyrokinetic predictions for momentum transport assessment also in TEM-dominated regimes. These results suggest that residual stress plays an important role in core rotation prediction for low-torque plasmas and should be accounted for in predictive models of future reactor scenarios.

\end{abstract}

\section{Introduction}
\label{sec:intro}

Achieving controlled nuclear fusion presents significant scientific and engineering challenges. To sustain fusion reactions, the plasma fuel must be heated to temperatures exceeding hundreds of millions of Kelvin and confined effectively. Tokamaks, such as ASDEX Upgrade (AUG) or DIII-D, are torus-shaped devices that use strong magnetic fields to confine and control this high-temperature plasma. One of the key quantities used to describe the plasma state is the toroidal rotation of the bulk plasma, as it influences the transport of plasma impurities \cite{Angioni_2012, casson2013, angioni2014tungsten, Casson_2015, Angioni2015} and can stabilize turbulence and influence confinement \cite{biglari1990influence, Hahm1994,  Waltz_1995_PoP, Hahm1995, Burrell1997, Terry2000}. Most importantly, plasma rotation can contribute to the mitigation and avoidance of harmful magnetohydrodynamic (MHD) events \cite{Strait1995, Garofalo2002, Reimerdes2007, Politzer2008, buttery2008, de_Vries_2011,Bardoczi_2024}, which can reduce performance or cause severe damage to fusion devices through disruptions.

Reliable prediction of rotation profiles in present and future reactors requires a detailed understanding of the sources, sinks, and transport of momentum in magnetically confined plasmas. Momentum transport within the tokamak core is influenced by several mechanisms, including diffusion, convection, and residual stress. Residual stress is a non-Fickian, off-diagonal contribution to the momentum transport \cite{Diamond2009, Camenen_PRL_2009, Camenen2011, Stoltzfus-Dueck_2012, Sun_2024} that generates an intrinsic torque capable of driving plasma rotation from rest \cite{Ida_1995, Rice_1998, Lee2003, Yoshida_2008_PRL, Yoshida_2009_PRL, Solomon2009}. Aside from the boundary condition of rotation at the pedestal top, the lack of validation of the theoretical understanding of momentum transport mechanisms remains the largest uncertainty in predicting core toroidal rotation profiles in future fusion devices.

Recent experimental and theoretical studies \cite{Kluy_2009,Camenen2011,Staebler_PRL_2013,Grierson_PRL_2017,Hornsby2018,Solomon2011,Angioni_PRL_2011,mcdermott2011effect, mcdermott2011core,Shi_2013,Na_2016,Shi_2017,Zimmmerann_PPCF_2026} have highlighted the role of turbulence dominating momentum transport in the plasma core. While dedicated research has focused on the ion-temperature-gradient (ITG) turbulence regime \cite{Zimmermann_2024} and ITG-trapped-electron-mode (TEM) mixed-mode regimes \cite{Zimmmerann_PPCF_2026}, thorough investigations of transport with dominant TEM are still lacking and only limited theoretical work in this regime by Kluy \textit{et al.} \cite{Kluy_2009} has been done. In addition, previous experimental approaches, either using balanced beams to cancel the rotation \cite{Solomon2009,Solomon2011} or being limited by the available beam configurations to high-torque scenarios \cite{Zimmermann_2024}, were not able to measure and gauge the effect of the background rotation and, consequently, $E \times B$ shearing on the transport processes. This is of high relevance, as future reactor devices will lack strong sources of external torque and, as such, the rotation level will likely be lower than in present-day devices. Therefore, to further complement those previous works, this study investigates momentum transport in TEM turbulence, focusing on the transition from ITG-dominated regimes to TEM regimes and examining the influence of modifications to the background $E \times B$ shearing by changing the background rotation level. 

The remainder of this paper is structured as follows. First, a short introduction to the applied methodology is given in Section \ref{sec:methodology}. Section~\ref{sec:reference} presents the reference discharge analysis, Section~\ref{sec:dataset} describes the full dataset and its turbulence characteristics, Section~\ref{sec:trends} discusses the parametric trends in the experimentally inferred transport coefficients, and Section~\ref{sec:GK_modeling} presents the comparison to linear gyrokinetic modeling with CGYRO. Finally, Section~\ref{sec:summary} summarizes the main findings and outlines directions for future work.

\section{Methodology}
\label{sec:methodology}

The analysis builds upon the momentum transport analysis framework previously developed for AUG \cite{Zimmermann2022, Zimmermann_NF_Letter,Zimmermann_NF_Isotope}, which has recently been ported and adapted for the DIII-D tokamak. The following description of the methodology is therefore kept very similar to that in Refs. \cite{Zimmermann_2024,Zimmmerann_PPCF_2026} and the reader is referred to Refs. \cite{Zimmermann_NF_Letter,Zimmermann_NF_Isotope} for a more detailed discussion.

In general, the framework enables separation of diffusive, convective, and residual stress contributions to core momentum transport through the analysis of experiments employing modulated neutral beam injection (NBI) heating. In this scenario, the NBI is used to produce controlled torque perturbations. Due to the availability of counter-current beam injection on the DIII-D tokamak, it is possible, for a subset of the studied experiments in this work, to produce a torque modulation without modulation of the injected NBI power. The resulting plasma rotation response can be measured using main ion and impurity ion charge exchange recombination spectroscopy (CER), see Refs. \cite{grierson2012active,chrystal2016improved} and references therein. In this work, rotation and ion temperature measurements were deduced from measurements of the carbon impurity, which were found to be in good agreement with residual boron measurements and are assumed to be representative for the bulk plasma rotation, see also the discussion in Appendix \ref{sec:appendix_main}.

The advantage of using the torque perturbation technique is that the measured rotation signal can be decomposed with Fourier analysis, allowing the different momentum transport contributions to be distinguished through their characteristic temporal behavior \cite{Zimmermann_NF_Letter}. It was found previously that the diffusive momentum fluxes are mainly correlated to the phase profiles, while the convective fluxes strongly impact the amplitude profiles \cite{Zimmermann2022}. The modeling framework itself is based on the toroidal momentum conservation equation introduced in \cite{fable2015toroidal} and solved using the ASTRA transport code \cite{ASTRA_Reference_Paper,Fable_2013,Tardini_2026}. The analysis is restricted to the plasma core by prescribing an experimental boundary condition for the toroidal rotation $v_\varphi$ at a location slightly inside the pedestal top. This avoids complications associated with edge phenomena such as edge-localized modes and is also adapted to the availability of edge CER. The sensitivity of the results to the precise choice of the boundary position is found to be small. Additionally, the influence of sawtooth (ST) events is minimized by restricting the analysis window to radii outside of those affected by ST ($\rho_\varphi\approx0.2$ for most discharges), as confirmed by ECE measurements. The flux-surface-averaged radial flux of toroidal momentum can be expressed as
\begin{equation}
    \Gamma_\varphi = - m_i n_i R \left(\chi_\varphi\,\frac{\partial v_\varphi}{\partial r} - V_\mathrm{c}\, v_\varphi\right)+\Pi_\text{RS} \text{,}
    \label{eq:momentum_flux}
\end{equation}
where $m_i$ is the main ion mass, $R$ is the flux-surface-averaged major radius, $\chi_\varphi$ represents the momentum diffusivity, and $v_\varphi=\langle R\Omega_\varphi\rangle$ denotes the toroidal velocity, with $\Omega_\varphi$ the toroidal angular frequency and $\langle\cdot\rangle$ describing a flux-surface average. The radial derivative $\partial/\partial r$ is taken with respect to the flux-surface-averaged minor radius $r$. The parameter $V_c$ denotes the convective velocity. The dimensionless quantity $-R V_c/\chi_\varphi$ is commonly referred to as the pinch number \cite{Peeters2007_PRL}. $\Pi_\text{RS}$ represents the residual stress flux, which corresponds to an intrinsic torque $\tau_\text{int} = -\partial V / \partial r \,\Pi_\text{RS}$, where $V$ is the plasma volume enclosed by the considered flux surface. Under the plasma core conditions studied here, poloidal rotation is insignificant when determining the average $v_\varphi$, and it is assumed to also be insignificantly altered by the modulated NBI.

In the present analysis, the momentum diffusivity $\chi_\varphi$ is parameterized as a linear function $a(\rho_\varphi)\cdot\chi_i$ of the normalized toroidal flux radius $\rho_\varphi$ and scales with the experimentally determined ion heat diffusivity $\chi_i$. This defines a spatially dependent Prandtl number $\mathrm{Pr} = \chi_\varphi / \chi_i=a(\rho_\varphi)$, which is typically on the order of unity in tokamak plasmas \cite{Peeters2005, waltz2007coupled, weiland2009symmetry, peeters2006toroidal, Strintzi2008, Peeters2011}. The ion heat diffusivity $\chi_i$ itself is obtained from power balance calculations based on experimental measurements and is given by $\chi_i=-Q_i/(n_i \nabla T_i)$, where $Q_i$ denotes the ion heat flux, $n_i$ the ion density, and $T_i$ the ion temperature. The radial profiles of both the convective velocity and the residual stress flux are represented using cubic polynomials, $b(\rho_\varphi)$ and $c(\rho_\varphi)$, constrained to remain continuous at $\rho_\varphi=0$ and linked to the momentum diffusivity to scale with the turbulence amplitude, as discussed in more detail in Ref. \cite{Zimmermann_NF_Letter}. 

Determination of the transport coefficients is performed using a statistical global optimization method~\cite{Storn1997}. In this procedure, the coefficients of the polynomials describing $a(\rho_\varphi)$, $b(\rho_\varphi)$, and $c(\rho_\varphi)$, scaling $\chi_\varphi$, $V_c$, and $\Pi_\text{RS}$, are iteratively adjusted to achieve the best agreement with the experimentally measured rotation profiles. The method also provides statistically robust estimates of the uncertainties in both the fitted coefficients and the modeled rotation profiles. These uncertainties reflect the experimental uncertainties in the input profiles. Finally, the external torque and heat fluxes arising from neutral beam injection are evaluated using the NUBEAM Monte Carlo code \cite{pankin2004}, which is part of the TRANSP analysis suite \cite{TRANSP_Reference_Paper, hawryluk1981empirical}. The heat fluxes from electron cyclotron heating (ECH) are calculated via the GENRAY code \cite{harvey2001genray}. These torque and heat fluxes are included self-consistently when solving the momentum conservation and power balance equations.

\section{Reference Discharge and Momentum Transport Analysis}
\label{sec:reference}

\begin{figure}
    \centering
    \includegraphics[width=0.55\linewidth]{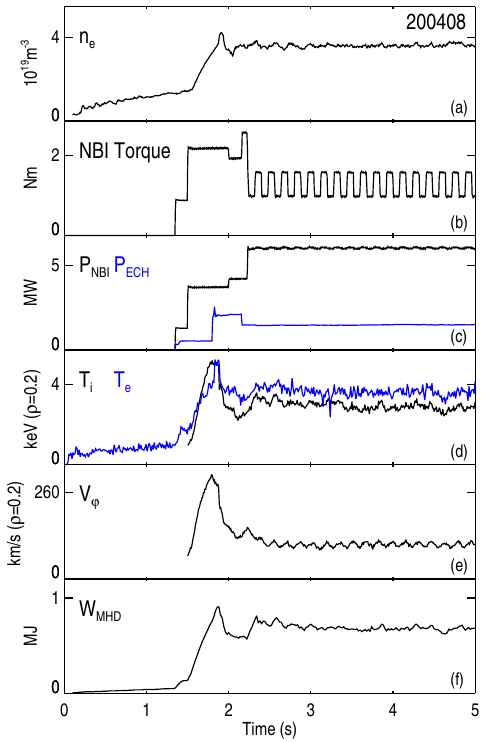}
    \caption{Main time traces of the reference discharge \#200408, the analysis time window is between $2.48$-$5.04$ s. Panel (a) shows the line-averaged electron density, which stabilizes after the ramp-up and L–H transition. Panel (b) displays the neutral beam power of the co- and counter-current injectors used to generate the 6.25 Hz torque modulation. Panel (c) shows the resulting total injected neutral beam power (black) together with the constant ECH power (blue). Panel (d) presents the central ion and electron temperatures at $\rho_\varphi=0.2$, which remain largely unaffected by the modulation (relative amplitudes at 6.25 Hz of about 2\% for $T_i$ and 0.1\% for $T_e$). Panel (e) shows the toroidal plasma rotation at $\rho_\varphi=0.2$, where the applied torque perturbation produces a clear periodic response (about 7\% amplitude at 6.25 Hz) used for the transport analysis. Panel (f) displays the stored magnetic energy, which remains constant (modulation of about 0.6\% at 6.25 Hz), indicating no significant changes in the magnetic equilibrium during the analyzed time interval. The modulations of $T_i$, $T_e$, $W_\mathrm{MHD}$, and $n_e$ are clearly smaller than the background fluctuations and the measurement noise of the respective time traces.}
    \label{fig:reference_timetraces}
\end{figure}

For this work, the momentum transport analysis framework from AUG \cite{Zimmermann2022,Zimmermann_NF_Letter} was ported to the DIII-D tokamak and a corresponding analysis workflow was established. Plasma discharge \#200408 was chosen as a reference discharge, as it constitutes an average, representative discharge within the studied dataset. In more detail, and as for all discharges featured in this work, it has a toroidal magnetic field of $B_t=-2$ T and a flat-top plasma current of $I_p=0.85$ MA, resulting in an edge safety factor of $q_\text{95}\approx5.8$. The effective charge number is $Z_\text{eff}\approx1.7$, as calculated from the carbon density assessed by CER and confirmed by Bremsstrahlung measurement, and is consistently and time-dependently included in the modeling. While the typical discharge time is around $5.5$s, the usable analysis time frame spans around $2.5$ s, e.g. from $2.48-5.04$ s in \#200408. The reason for shortening the analysis time frame becomes obvious when looking at the main time traces presented in Fig. \ref{fig:reference_timetraces}, showing a stabilization of the plasma density (Panel a) and temperatures (Panel d) after the ramp-up phase and the L-H transition. The line-averaged core density, for all but the \textit{deep TEM} discharges, is around $3.7\cdot10^{19}$ m${}^{-3}$. As shown in Panel (b), a combination of co- and counter-current beams was used to induce a significant torque perturbation with a 6.25 Hz frequency (about $1.1$ Nm amplitude). As shown by the black curve in Panel (c), the co-counter-balanced injection scheme results in a negligible variation of the injected neutral beam power. ECH is at constant power with $\approx 0.5$ MW aimed at $\rho_\varphi\approx 0.17$ and $1$ MW aimed at $\rho_\varphi\approx 0.38$ resulting in a total power of $1.5$ MW, see blue curve.

The applied heating scheme translates to non-perturbed temperature time traces, see the black curve for $T_i$ and the blue curve for $T_e$ in Panel (d). The modulation results in a clearly visible perturbation of the toroidal plasma rotation (about 10\% in amplitude), shown in Panel (e), which will later be used for the analysis. No major variations of the magnetic topology were observed, and the stored magnetic energy is rather constant, as shown in Panel (f). While all main background quantities are kept constant during the analyzed time window, they are all consistently included at full time dependence in the following modeling. As the density is unperturbed, variations in the plasma angular momentum \(L_\varphi = m_i n_i R v_\varphi\) arise mainly from changes in the toroidal rotation velocity. For this reason, the model–experiment optimization is performed by fitting the rotation velocity rather than the angular momentum itself, even though angular momentum is the quantity that is physically conserved. In addition, it is well justified in the remainder of this work to refer to time-averaged quantities when describing the plasma conditions for each experiment.

While the co-counter-balanced modulation at constant injection power requires a certain minimum NBI power to be applied, some of the discharges in this work required overall low NBI powers (deep TEM) or strong counter-current injection to minimize the background rotation, see Table \ref{tab:shotlist}. For such limitations to the scenario, some of the analyzed discharges feature only co-current torque perturbation, as done, e.g., in discharge \#200419. In that discharge, this results in roughly half of the torque perturbation (about $0.43$ Nm), with a modulation of the injected power of about $0.6$ MW (similar to previous work on AUG, see Ref. \cite{Zimmermann_NF_Isotope}), with a relative rotation modulation of 30\% and density modulation below 1\%. Temperature and stored energy modulations are smaller than 10\%, as obtained from a Fourier analysis at the driven frequency. The latter are comparable to or smaller than the noise on the measurement time traces, i.e., the scatter of the signals caused by imperfections of the measurements and by background variations such as ELMs and ST. These values are well within the range for which this methodology has previously been validated \cite{Zimmermann2022} and applied \cite{Zimmermann_NF_Letter,Zimmermann_NF_Isotope,Zimmermann_2024,Zimmmerann_PPCF_2026}, and possible modulations of the background turbulence amplitude due to the heating perturbation are accounted for by scaling all transport coefficients with the experimentally measured ion heat diffusivity.

\begin{table}
\centering
\begin{tabular}{lllllllll}
\hline
Discharge & Time & Mod.& $P_\text{NBI}$ & $\tau_\text{NBI}$& $P_\text{ECH}$ & ECH& $n_\text{e} $ & $Z_\text{eff}$ \\
\#200...& [s] &  scenario& [MW] &  [Nm] & [MW] & max. $\rho_\varphi$ & [$10^{19}$ m$^{-3}$] & core avg. \\
\hline
'403 & 2.64--4.88 & co-ctr. & 5.9 & med. 1.3 & low 0.6 & 0.21 & 3.7 & 1.7\\
'406 & 2.48--5.04 &co-ctr. & 5.9 & med. 1.3 & max. 3.2 & 0.40 & 3.7 & 1.7\\
'408 & 2.48--5.04 &co-ctr. & 5.9 & med. 1.3 & med. 1.5 & 0.20 & 3.7 & 1.7\\
'414 & 2.96--5.20 & co& 5.0 & low -0.6 & max. 3.2 & 0.40 & 3.7 & 1.6\\
'416 (TEM) & 2.48--4.40 & co& 2.2 & med. 1.9 & max. 3.2 & 0.39 & 2.7 & 1.7\\
'418 (TEM) & 2.48--4.40 & co& 2.2 & med. 1.9 & max. 3.2 & 0.22 & 2.7 & 1.7\\
'419 & 3.76--5.20 & co& 5.0 & low -0.6 & low 0.6 & 0.17 & 3.7 & 1.9 \\
\hline
\end{tabular}
\caption{Summary of experimental scenarios and main parameters. The injected powers and the NBI torque are time-averaged values, where the NBI torque is the total torque integrated over the entire plasma volume from the magnetic axis to the last closed flux surface. The column ``ECH max.'' gives the radial location of the maximum of the time-averaged, gyrotron-summed ECH power deposition profile from TORBEAM in TRANSP. The power and deposition location of each individual gyrotron are shown in Appendix~\ref{app:ech}.}
\label{tab:shotlist}
\end{table}

\begin{figure}
    \centering
    \begin{overpic}[width=1.0\linewidth]% remove grid,tics once placed
        {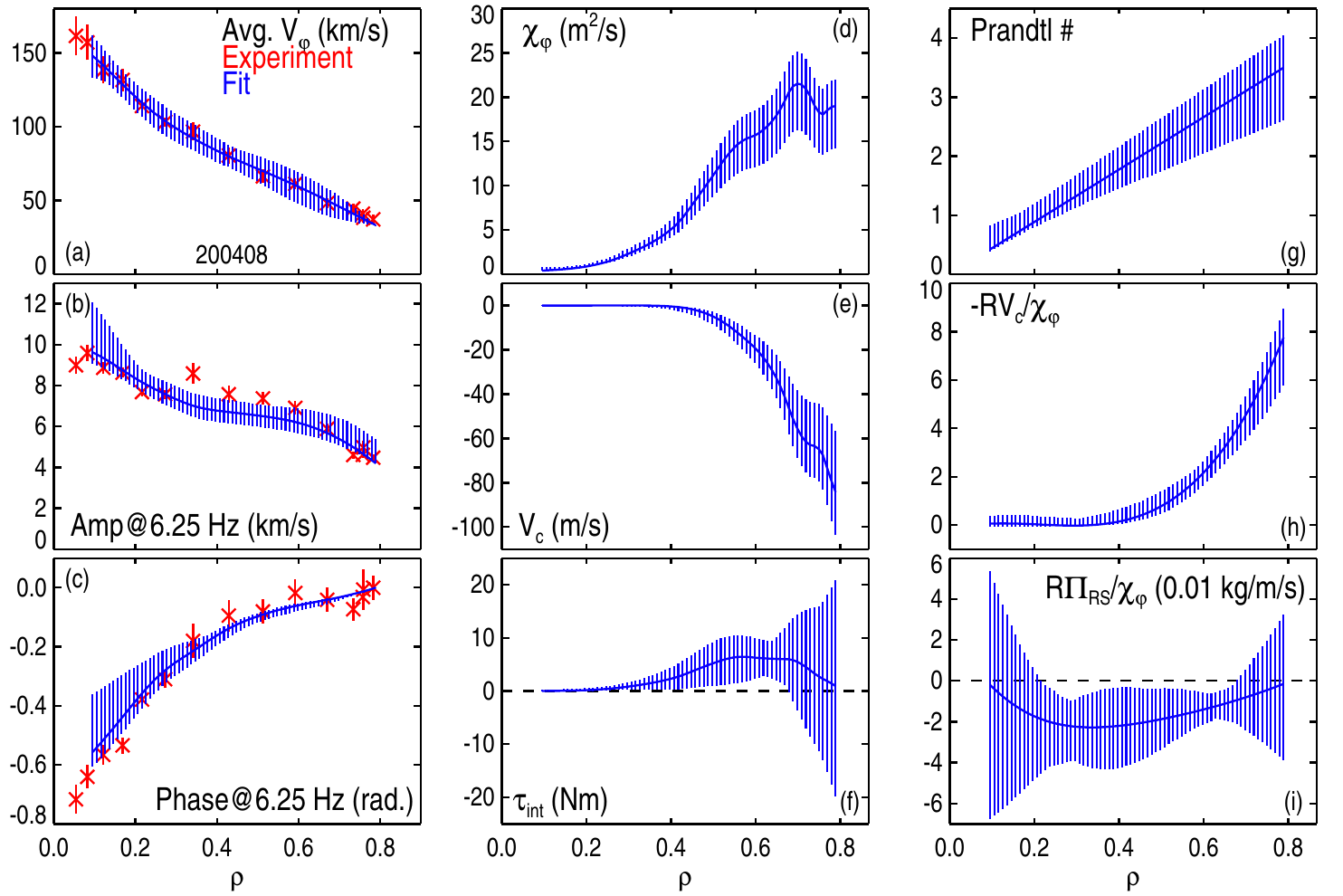}
        \put(39,18){\sffamily co-current}
        \put(39,11){\sffamily ctr.-current}
        \put(80,7){\sffamily co-current}
        \put(80,18){\sffamily ctr.-current}
    \end{overpic}
    \caption{Results of the momentum transport analysis for the selected reference discharge \# 200408. Panels (a)--(c) compare experimental measurements (red markers with error bars) with simulated profiles (blue lines with error bars): (a) steady-state toroidal rotation profile, (b) perturbation amplitude, and (c) phase profile for the modulation frequency. Panels (d)--(i) show the inferred transport quantities: (d) momentum diffusivity $\chi_\varphi$, (e) convective velocity $V_c$, (f) intrinsic torque $\tau_\text{int}$, (g) Prandtl number, (h) pinch number, and (i) residual stress flux normalized by $\chi_\varphi$. Error bars on the experimental data result from statistical uncertainties of the experimental analysis; uncertainties on the fitting results represent a $1\sigma$ environment around the solution.}
    \label{fig:reference_analysis}
\end{figure}
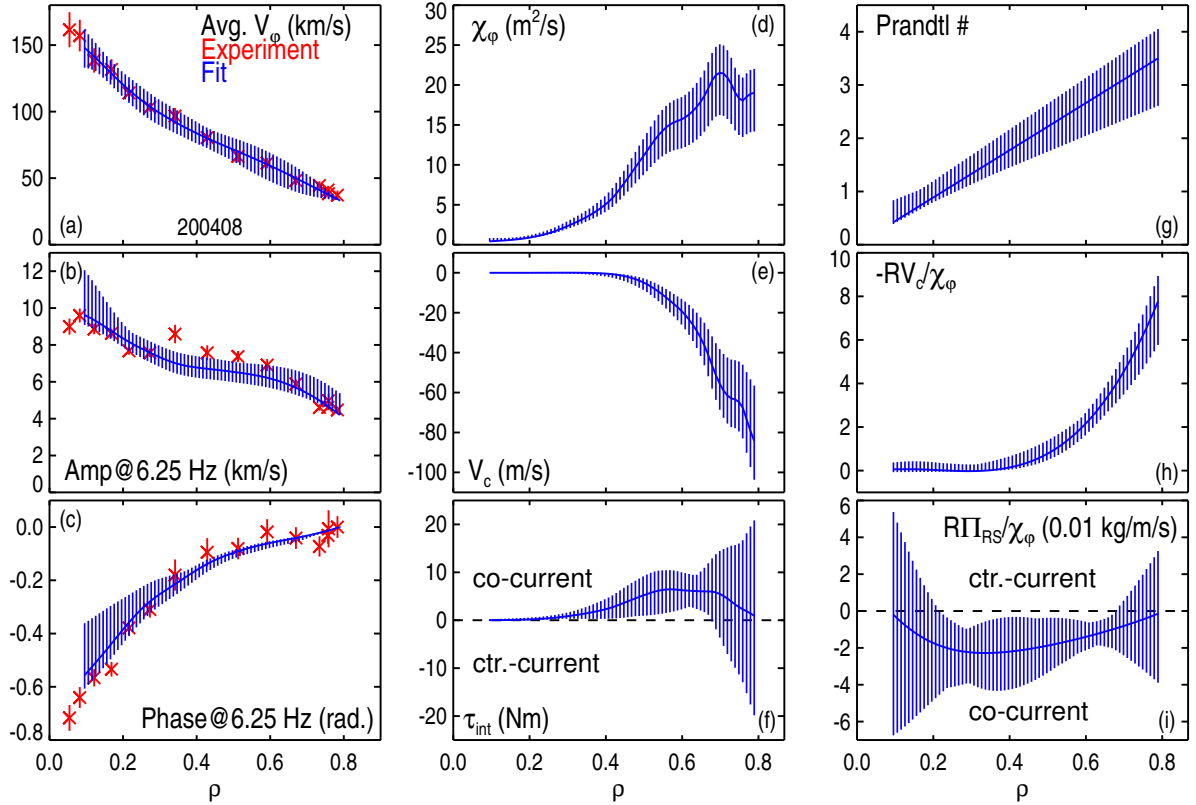

The previously established momentum transport analysis was applied to all studied discharges. Figure~\ref{fig:reference_analysis} presents the results for the selected reference discharge. As already noted, the analysis range was adapted for all discharges to provide smooth edge boundary conditions. The inner boundary of the analysis was set to roughly $\rho_\varphi=0.2$, depending on the exact radial positions of the CER channels, to prevent near-axis MHD from having a significant effect on the analysis. This radius is an approximate average of the ST inversion radius (for discharges with ST) and the reconstructed location of the $q=1$ surface (for discharges without). All key parametric dependencies are evaluated at $\rho_\varphi=0.4$ and $0.6$, well outside of this region, and reasonable variations of the inner boundary were verified to have no significant effect on the results. The simulated profiles (blue lines with error bars) closely match the experimental measurements (red markers with error bars from the fitting of the CER spectra), accurately reproducing the steady-state rotation profile shown in Panel~(a). The perturbation amplitude in Panel~(b) is generally well reproduced. The overall phase shift, shown in Panel~(c), is also well matched. The modeled time evolution at mid-radius, not shown for brevity, exhibits no evidence of drifts, indicating that the reconstructed transport coefficients provide a self-consistent steady-state solution. All uncertainties reflect a variation of $1\sigma$ around the solution, as discussed in Ref.~\cite{Zimmermann_NF_Isotope}. The error bars can appear somewhat asymmetric for the Prandtl numbers, as the prescribed linear shape of $Pr$ is constrained by the boundary condition requiring it to remain positive over the entire plasma radius.

The same figure also summarizes the inferred transport coefficients. Panel~(d) shows the resulting momentum diffusivity profile, which increases with radius and reaches relatively high values. Panel~(g) shows $Pr$, which increases with radius up to values of approximately $3.5$. Panel~(e) shows the convective velocity, which is obtained by multiplying the pinch number in Panel~(h) with the momentum diffusivity $\chi_\varphi$. The pinch number itself is very flat and small in the inner core and increases toward the edge of the analysis domain. It should be noted that the modeled pinch numbers include contributions from Coriolis pinch and particle transport. However, the particle convection momentum flux was estimated in TRANSP and found to be negligible.

Panel~(f) shows the inferred intrinsic torque values. Although the experimental uncertainties are large, a cumulative co-current, positive intrinsic torque is observed, increasing toward the edge of the analysis domain before the uncertainties become too large to yield meaningful results. The shape and magnitude are very similar to those reported in previous investigations on AUG and \diiid{} ~\cite{Solomon2009, Solomon2011, Zimmermann_NF_Letter, Zimmermann_NF_Isotope}. The corresponding residual stress fluxes, normalized by $\chi_\varphi$, are shown in Panel~(i). Note the sign convention, in which a negative flux corresponds to an inward, co-current flux. As for the pinch number, normalization by $1/\chi_\varphi$ is considered most useful for the heating scans performed here. Overall, these plots also illustrate why, in the remainder of this work, most parametric dependencies are evaluated at $\rho_\varphi = 0.6$, where the uncertainties for most studied quantities remain sufficiently small to allow meaningful conclusions to be drawn. As the total momentum flux is often a balance between the diffusion, convection, and intrinsic fluxes, results shown later will use the quantities normalized by the diffusivity in order to make the relevant trends clear.

\section{Experimental Dataset and Turbulence Regimes}
\label{sec:dataset}

\begin{figure}
    \centering
    \includegraphics[width=0.4\linewidth]{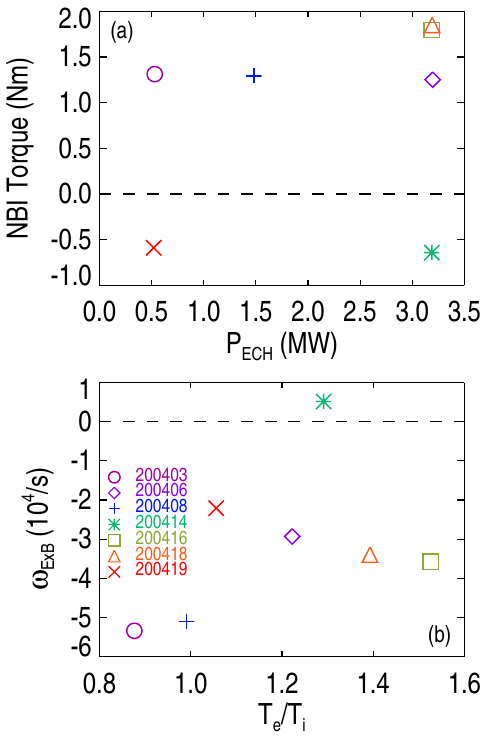}
    \caption{Key parameters of the studied dataset. Panel (a) shows main engineering parameters of the applied NBI torque, integrated over the entire plasma volume from the magnetic axis to the last closed flux surface, and injected ECH power. Panel (b) shows the obtained variation in $E\times B$ shearing and $T_e/T_i$ ratio at $\rho_\varphi=0.6$. The lower bound of the turbulence growth rates from gyrokinetic modeling, roughly $4\times10^{4}\,\mathrm{s^{-1}}$, is larger than most of the shearing rates shown (see the discussion later in Sec.~\ref{sec:dataset}).}
    \label{fig:parameter_range}
\end{figure}

The main scope of the conducted experiments (see the overview in Table \ref{tab:shotlist}) was to advance momentum transport studies toward reactor-relevant conditions. This implies investigating turbulent momentum transport in a plasma scenario with, first, dominant electron heating and, second, low-torque and, consequently, low-rotation regimes, where $E \times B$ shearing rates are near or below turbulence growth rates. Both were achieved by varying the mixture of NBI ($2$--$6\, \mathrm{MW}$) and ECH heating (up to $\approx3\,\mathrm{MW}$), while also relying on \diiid{}'s combination of co- and counter-current NBI to minimize input torque. The spanned parameter range, in terms of engineering parameters, is shown in Fig. \ref{fig:parameter_range}(a). One can see that the total applied torque ranges from $-0.5$--$2.0$ Nm with a simultaneous scan in ECH power between $0.5$--$3.2$ MW. This translates into the physics parameters shown in panel (b), namely a variation of the electron-to-ion temperature ratio from $0.9$--$1.55$ and a substantial variation of the $E\times B$ shearing rate, which is determined from experimentally measured contributions to the force balance \cite{Burrell1997}.

In more detail, in the scanned parameter space, there are certain axes of scans: \#200403, '408, and '406 feature strong $P_\text{NBI}=5.9$ MW with constant input torque $\tau_\text{NBI}$ and a scan in ECH power from $0.6, 1.45, 3.2$ MW. With respect to the high NBI power and high density, this likely represents a scan from an ITG to an ITG-TEM mixed-mode regime. Discharges '414 and '419 are low rotation cases with $\tau_\text{NBI}\approx-0.6$ Nm and roughly match the input heating of \#200406 and '403, showing the effect of the background $E\times B$ shearing and constituting a heating scan at low input torque. Finally, a major aim of these experiments was to combine a deeply TEM-dominated turbulence regime with the beam modulation technique. To this end, discharges \#200416 and '418 were designed to obtain the deepest possible TEM (given the constraints of continuous core rotation measurements and the beam modulation). They exhibit $P_\text{ECH}\approx3.2$ MW versus $P_\text{NBI}\approx 2.2$ MW and a very low core averaged density of $2.7\cdot10^{19}$ m${}^{-3}$ which increases beam penetration, but is beneficial for accessing TEM by reducing the collisions that de-trap electrons.

\begin{figure}
    \centering
    \begin{overpic}[width=0.4\linewidth]% remove grid,tics once placed
        {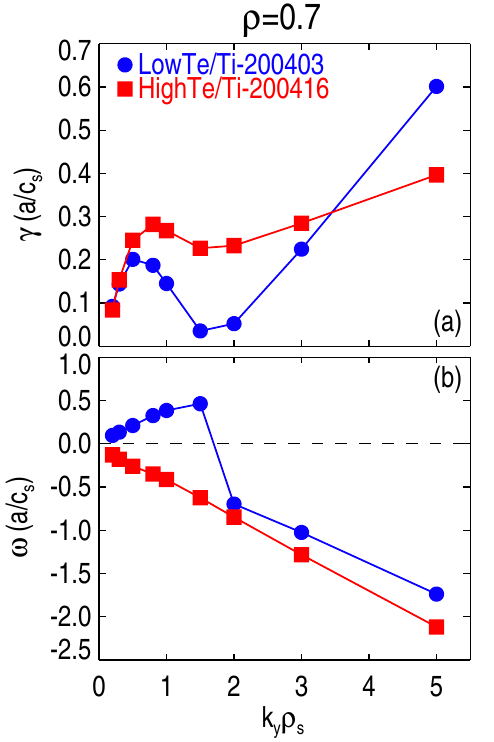}
        \put(35,45){\sffamily ion direction}
        \put(17,15){\sffamily electron direction}
    \end{overpic}
    \caption{Linear CGYRO growth rates (Panel a) and real frequencies (Panel b) from local flux-tube simulations at $\rho_\varphi\approx0.7$ for discharges \#200403 (ITG case), and \#200416 (TEM case) as a function of the tested wave number $k_y\rho_s$.}
    \label{fig:cgyro_turbulence_regime}
\end{figure}

A first estimate of the success in reaching a TEM-dominated turbulence regime can be obtained using gyrokinetic calculations. Simulations were run using linear CGYRO to characterize dominant instabilities and, later, in Section \ref{sec:GK_modeling}, to isolate the diffusive and pinch components of momentum flux in the plasma bulk region. CGYRO is an Eulerian, flux-tube gyrokinetic code which calculates perturbations to the plasma distribution with an assumed Maxwellian background\cite{Belli_2017}. The simulations presented used input from the experimental magnetic equilibrium and species-dependent temperature and density profiles. Electrostatic and transverse electromagnetic fluctuations were considered, and three kinetic species (deuterium, electrons, and a carbon impurity) were included. Perpendicular wavenumbers $k_y\rho_{s,D}$ ranged from $0.05 – 10.00$. Figure \ref{fig:cgyro_turbulence_regime} shows the result of such a scan at $\rho_\varphi\approx0.7$ for discharges \#200403 (considered deep ITG, $P_\text{NBI}=5.9$ MW, $P_\text{ECH}=0.6$ MW) and \#200416 (considered deep TEM, $P_\text{NBI}=2.2$ MW, $P_\text{ECH}=3.2$ MW). Panel (a) shows a typical spectrum with small wave numbers $k_y \rho_s$ (x-axis), non-zero growth rates (y-axis), and a typical increase in growth rates for higher wave numbers. Panel (b) shows the corresponding dominant wave frequency, with negative numbers corresponding to mode propagation in the electron diamagnetic direction and positive numbers corresponding to mode propagation in the ion diamagnetic direction. It should be noted that usually the low $k_y \rho_s$ contribute most strongly to the transport, therefore confirming dominant ITG and TEM conditions in each of the studied discharges. To further examine this, for $\rho_\varphi=0.7$ and $k_y \rho_s=0.5$, the input gradients $R/L_{n_e}$, $R/L_{T_i}$, and $R/L_{T_e}$ of the simulation were individually varied by a factor of 0.8--1.2 around their experimental values and the effect on the resulting growth rates was tracked. This demonstrated that the modes in \#200403 are most strongly destabilized by $R/L_{T_i}$, in agreement with the theoretical picture of ITG. In \#200416, the mode is destabilized by $R/L_{T_e}$, i.e. likely an electron temperature gradient driven TEM in the intermediate $k_y \rho_s$ range. In principle, as shown by the strongly negative values for $k_y \rho_s \approx 5$, also electron-temperature-gradient (ETG) driven turbulence could be present. These effects, on the order of the electron gyro-radius, cannot meaningfully impact the momentum transport due to the much larger gyro-radii of the ions. It should be noted, however, that strong ETG could still drive electron heat transport, affecting the overall power balance. Such more detailed considerations are left for future work.

\begin{figure}
    \centering
    \includegraphics[width=0.85\linewidth]{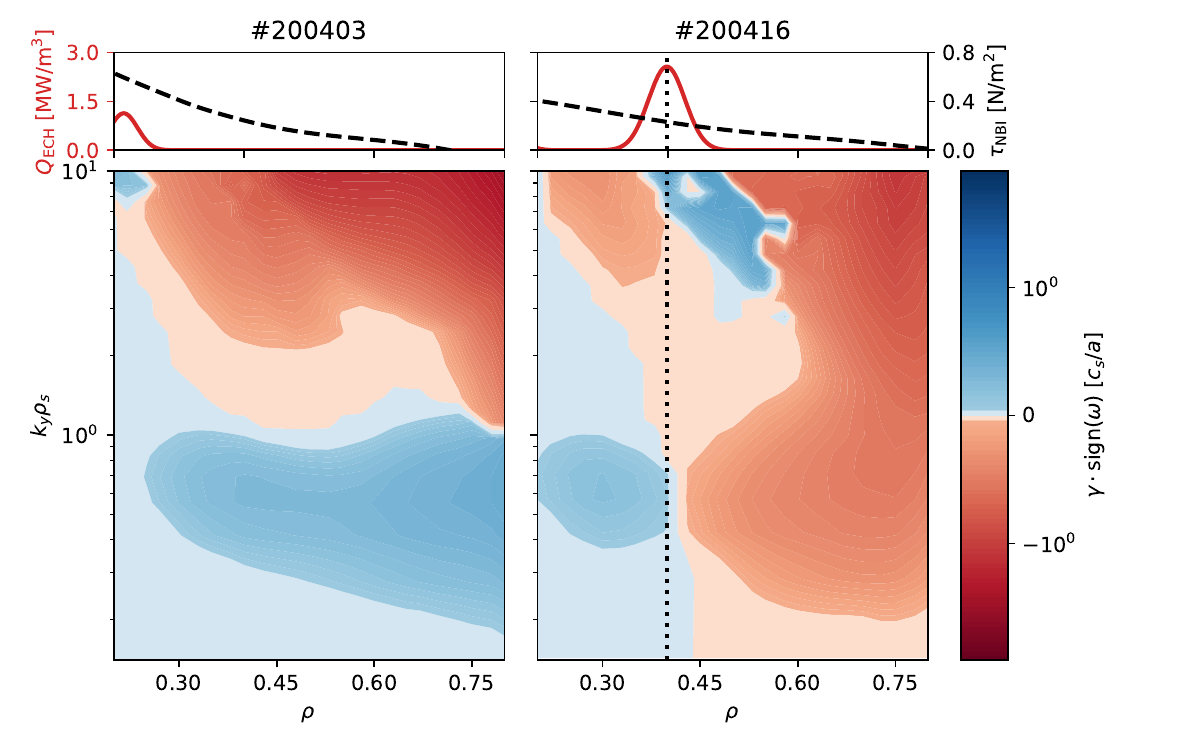}
    \caption{TGLF calculations for the ITG case \#200403 (left) and the TEM case \#200416 (right). Colors indicate the dominant instability (red: electron diamagnetic direction, blue: ion diamagnetic direction) as a function of radius (x-axis) and binormal wavenumber (y-axis). Insets above show the ECH power density (red solid, left axis) and the NBI torque density (black dashed, right axis) from TRANSP, averaged over the analysis phase. The dotted line marks the off-axis ECH deposition location in the TEM case.}
    \label{fig:tglf_turbulence_regime}
\end{figure}

These insights can be deepened by applying the gyrofluid TGLF code \cite{Staebler_2007} to both cases, which is computationally cheaper and allows for tracking sub-dominant modes. For this study, experimental input (time averaged over the analysis phase) was used for TGLF calculations with the SAT2 saturation rule. The results are shown in Fig. \ref{fig:tglf_turbulence_regime}, with the ITG case \#200403 on the l.h.s. and the TEM case \#200416 on the r.h.s. The color plots show the growth rates (coloring) over radius (x-axis) and mode number spectra (y-axis). Deep red areas indicate TEM-dominated turbulence, while blue areas indicate ITG-dominated turbulence. Looking at the vertical cross-section at $\rho_\varphi\approx0.7$, the same ITG dominance is found for \#200403 for the lowest $k_y \rho_s$, switching to TEM for higher wave numbers, consistent with the CGYRO results. This holds for the entire analysis range of the transport analysis, from $\rho_\varphi=0.2-0.8$. For the deep TEM case on the r.h.s., however, strongly TEM-dominated turbulence is found for radii outside of $\rho_\varphi = 0.4$, with some weak ITG contributions around mid-radius. The small plots above the color plots show the deposited power density of the ECH heating, which is small and on-axis for \#200403 and much stronger and off-axis in \#200416. The vertical dotted line on the r.h.s. demonstrates that outside of the ECH deposition, a pronounced transition to TEM is observed, while the ECH is not able to significantly modify the nature of the turbulence inside the deposition region. The sub-dominant modes were found to be pronounced in the TEM direction for \#200416 and weak, fluctuating between ITG and TEM, for \#200403. The second TEM discharge \#200418 shows very similar results and is not shown here for brevity. In summary, the TGLF calculations also support the assumption that the parameter scan reaches a TEM-dominated turbulence regime.

Additional experimental insight into the turbulence characteristics can be obtained from Beam Emission Spectroscopy (BES) \cite{BES_reference} and Doppler Backscattering (DBS) measurements \cite{peebles2010novel}. While these diagnostics do not provide a direct characterization of the underlying instability, they allow examination of fluctuation amplitudes and propagation properties, which can be compared with the turbulence regime suggested by the gyrokinetic and gyrofluid modeling. It should be noted that, unfortunately, due to a secondary experiment, the radial coverage of these diagnostics was consciously placed outside the analysis range considered here. The BES array covered approximately $\rho_\varphi\approx0.7$--$1.0$, depending on the configuration. The analysis focused on density fluctuation levels inferred from measured intensity fluctuations in the frequency band $20$--$150$ kHz during ELM-free intervals. An increase in fluctuation amplitude is observed for the higher-ECH case (\#200406, $P_\text{ECH}=3.2$ MW, $P_\text{NBI}=5.9$ MW) compared to the lower-ECH ITG discharge (\#200403, $P_\text{ECH}=0.6$ MW, $P_\text{NBI}=5.9$ MW), likely resulting from the overall higher heating power and, therefore, higher turbulence amplitude. Complementary information is obtained from DBS measurements, which probe density fluctuations at perpendicular wave numbers around $k_y \rho_s \sim 1$ at $\rho_\varphi\approx0.7$. A comparison between the low-ECH ITG case (\#200403, $P_\text{ECH}=0.6$ MW, $P_\text{NBI}=5.9$ MW) and the higher-ECH case (\#200406, $P_\text{ECH}=3.2$ MW, $P_\text{NBI}=5.9$ MW) indicates systematically larger normalized fluctuation levels in the higher-ECH discharge, consistent with the idea that higher overall heating power drives stronger turbulence and in agreement with the BES observations. The increase is particularly pronounced during inter-ELM periods, suggesting that the stronger electron heating enhances the drive of electron-scale turbulence. The measured cross-phase is negative for both \#200403 and \#200406, indicating fluctuation propagation in the electron diamagnetic direction in the laboratory frame. This observation is consistent with the presence of TEM-like turbulence shown in the TGLF scan in Fig. \ref{fig:tglf_turbulence_regime} around $k_y \rho_s\approx1$, but firm conclusions are difficult to draw as TGLF suggests the transition from ITG to TEM happening around that wavenumber in the spectra.

Finally, one of the central aspects of this experiment was to access plasma conditions in which the background $E\times B$ shearing rate becomes comparable to or smaller than the characteristic turbulence growth rates. Such conditions are expected to be more representative of future reactor devices, where external torque sources will be weak (relative to the angular momentum of such plasmas) and rotational shear will therefore play a reduced role in suppressing turbulence. An approximate estimate of the relevant growth-rate scale can be obtained from the gyrokinetic calculations shown in Fig.~\ref{fig:cgyro_turbulence_regime}(a). With $c_s\sim\sqrt{T_e/m_i}$ the ion sound speed normalization in CGYRO, the conversion factor $c_s/a$ is typically on the order of $5\times10^{5}\,\mathrm{s^{-1}}$ for the presented discharges, with rather small variations due to the temperature changes. The dimensionless growth rates obtained from the CGYRO spectra therefore correspond to physical values in the range of roughly $4\times10^{4}$--$1\times10^{5}\,\mathrm{s^{-1}}$ for the lowest $k_y\rho_s$ modes in discharge \#200403, with larger values occurring at higher wave numbers. Very similar growth-rates are obtained for the TEM-dominated discharge \#200416. When these values are compared with the experimentally inferred $E\times B$ shearing rates shown in Fig.~\ref{fig:parameter_range}(b), it becomes evident that the scanned parameter range indeed approaches conditions where the background shear is comparable to or even below the turbulence growth rates for at least one point of the dataset. 

\section{Observed Parametric Dependences}
\label{sec:trends}

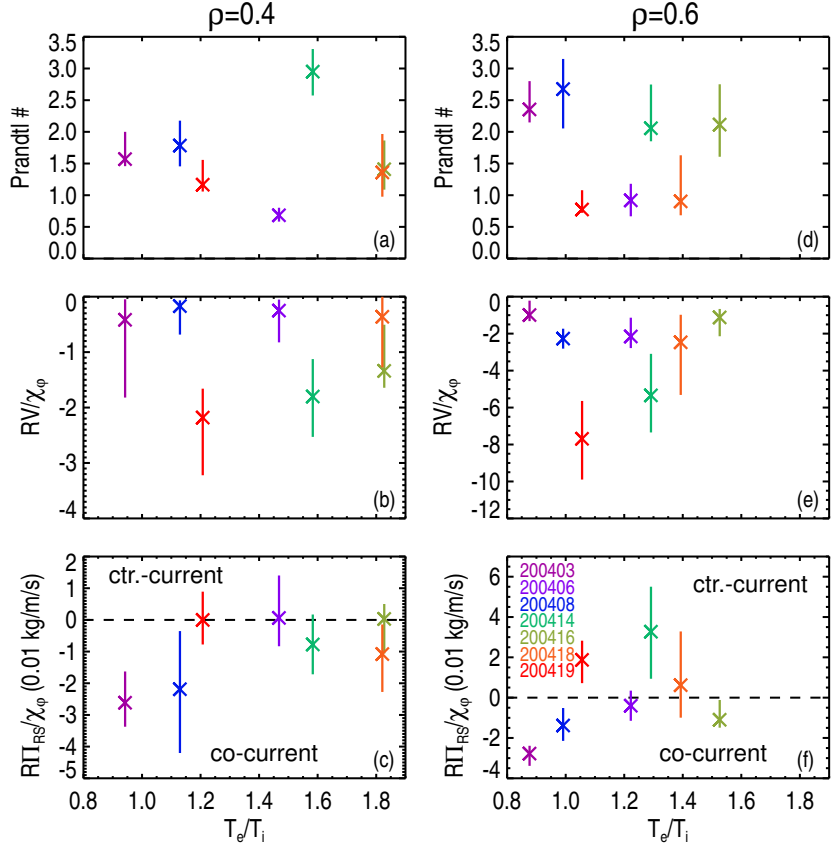
\begin{figure}
    \centering
    \begin{overpic}[width=0.7\linewidth]{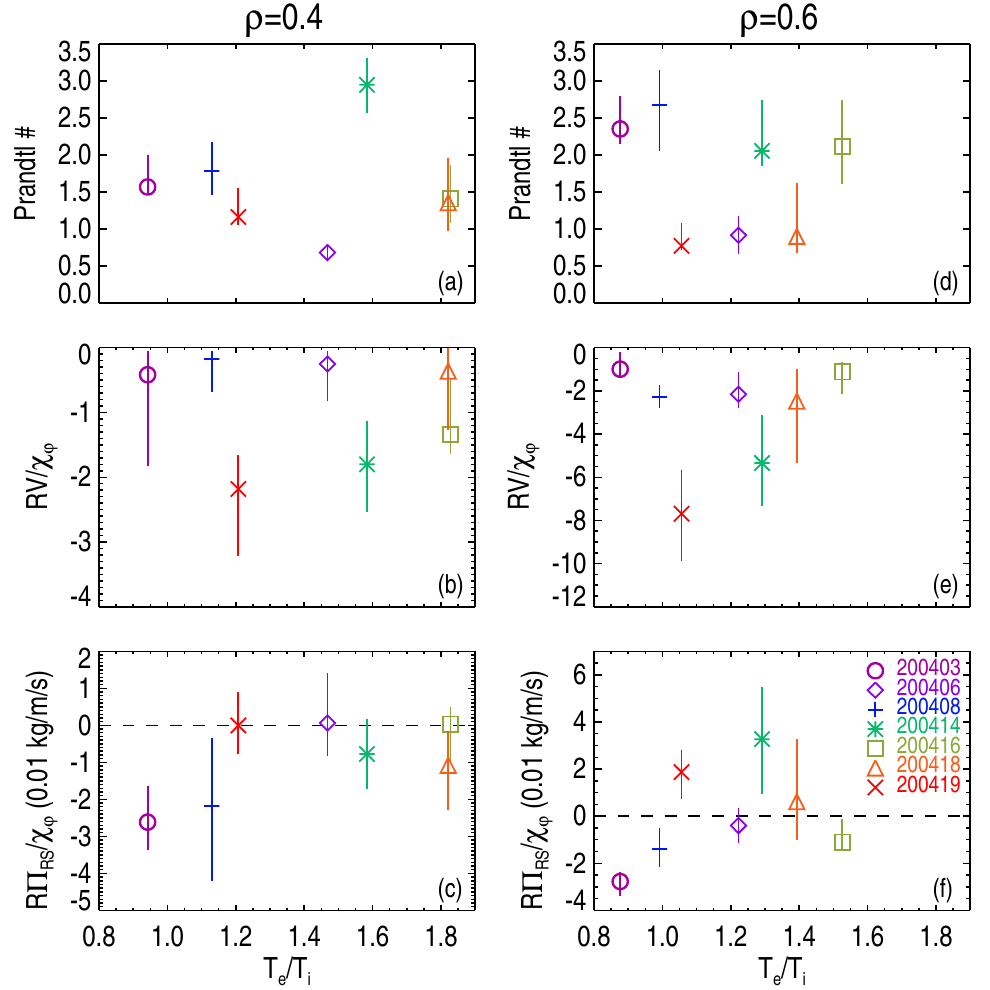}
    \put(13,31){\sffamily \small ctr.-current}
    \put(25,10){\sffamily \small co-current}
    \put(78,10){\sffamily \small co-current}
    \put(62,31){\sffamily \small ctr.-current}
    \end{overpic}
    \caption{Dependence of inferred transport parameters on the temperature ratio $T_e/T_i$, which orders the dataset from ITG-dominated to TEM-dominated conditions. Panels (a) and (d) show the Prandtl number at $\rho_\varphi=0.4$ and $\rho_\varphi=0.6$, respectively. Panels (b) and (e) show the corresponding pinch numbers, while Panels (c) and (f) present the residual stress flux normalized by $\chi_\varphi$. Each point represents a discharge from the dataset, with error bars reflecting the statistical uncertainties from the transport analysis.}
    \label{fig:TeTi_scan}
\end{figure}

The methodology discussed in Section~\ref{sec:methodology}, and illustrated for a single discharge in Section~\ref{sec:reference}, was applied to the entire dataset described in Section~\ref{sec:dataset}. All experimental steady-state amplitude and phase profiles were reasonably well matched. As shown, the dataset provides scans in background $E\times B$ shearing and spans turbulence regimes ranging from ITG- to TEM-dominated conditions. In this section, $T_e/T_i$ is used to approximately order the datapoints from ITG to TEM regimes. Higher electron temperature fractions typically lead to steeper electron temperature gradients, which favor TEM turbulence, while lower $T_e/T_i$ conditions tend to favor ITG-driven turbulence. In addition, within this dataset, $T_e/T_i$ acts as an approximately one-to-one proxy for the electron-to-ion heat flux ratio, which varies between $Q_e/Q_i = 0.6$-$1.8$ for $T_e/T_i = 0.8$-$1.4$ at $\rho_\varphi = 0.6$. The dependence of the transport parameters is examined at two radial locations in the plasma core, $\rho_\varphi = 0.4$ and $\rho_\varphi = 0.6$. The corresponding dependences of the Prandtl number, pinch number, and residual stress are shown in Fig.~\ref{fig:TeTi_scan}.

First, the Prandtl numbers are considered, shown at $\rho_\varphi = 0.4$ in Panel~(a) and at $\rho_\varphi = 0.6$ in Panel~(d). No clear trend in the Prandtl number is observed during the transition from ITG-dominated conditions (low $T_e/T_i$) to TEM-dominated conditions (high $T_e/T_i$). At $\rho_\varphi = 0.4$, most points overlap within their error bars, with the exception of discharges \#200406 and \#200414 in the intermediate $T_e/T_i$ range. At $\rho_\varphi = 0.6$, two groups of Prandtl numbers appear but with no particular dependence on $T_e/T_i$. Discharges \#200419, \#200406, and \#200418 exhibit lower values below unity, while another group clusters around values near $Pr \approx 2.5$. In general, only weak variations are expected from previous theory work due to the approximate invariance of the Prandtl number with respect to plasma conditions, as discussed in Refs.~\cite{Strintzi2008,Kluy_2009,Zimmermann_2024}. This may indicate that the underlying assumption behind the Prandtl number, namely that momentum diffusivity scales with ion heat diffusivity, is not a well-defined quantity in the transition regime between ITG and TEM turbulence, where some of the lower data points reside. However, attempts to establish such a correlation via statistical analysis were unsuccessful. In particular, there is no systematic deviation of the $Z_\text{eff}$ values of these lower $Pr$ cases, see Table \ref{tab:shotlist}. It should be noted that the analysis itself does not require the Prandtl number to be a well-defined quantity. The scaling with $\chi_i$ mainly accounts for possible modulations of the turbulence amplitude (see Section~\ref{sec:reference}), while the physically relevant quantity is the resulting momentum diffusivity. If the Prandtl number concept does not hold in the transition regime, the fitted value is simply an effective number to produce the required diffusivity, without affecting the other inferred quantities.

\begin{figure}
    \centering
    \settoheight{\figH}{\includegraphics{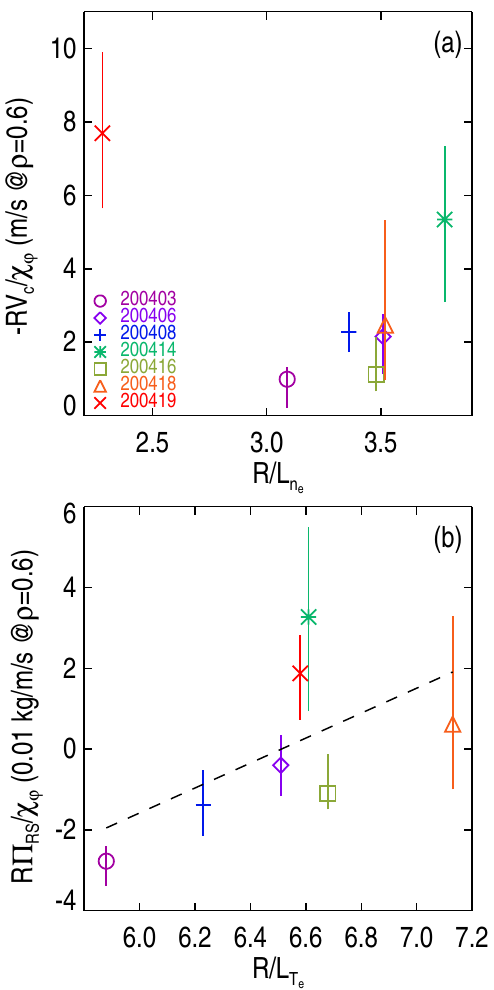}}% natural height of the stacked figure
    \setlength{\cutA}{0.497\figH}% lower part (panel b), removed for panel (a)
    \setlength{\cutB}{0.503\figH}% upper part (panel a), removed for panel (b)
    \edef\optA{width=0.4\linewidth,trim=0pt \the\cutA\space 0pt 0pt,clip}
    \edef\optB{width=0.4\linewidth,trim=0pt 0pt 0pt \the\cutB,clip}
    \begingroup\expandafter\overpic\expandafter[\optA]{Figure7_new.pdf}
    \endoverpic\endgroup\hspace{0.03\linewidth}
    \begingroup\expandafter\overpic\expandafter[\optB]{Figure7_new.pdf}
        \put(35,25){\sffamily co-current}
        \put(22,60){\sffamily ctr.-current}
    \endoverpic\endgroup
    \caption{Comparison of gradient scale lengths versus pinch and residual stress numbers at $\rho_\varphi=0.6$.}
    \label{fig:combined}
\end{figure}

The pinch numbers, shown in Panels~(b) and (e), also do not exhibit an explicit trend across the ITG–TEM transition. While most points overlap within their uncertainties, the low-rotation and low-$E\times B$ shearing cases \#200414 and \#200419 show significantly larger values. Around $\rho_\varphi = 0.6$, see  Fig.~\ref{fig:combined}(a), the logarithmic density gradient correlates best with the inferred pinch numbers, reminiscent of the Coriolis momentum pinch mechanism discussed in Refs.~\cite{Peeters2007_PRL,Tala_2009}, explaining the higher value for \#200414. In that plot, \#200419 is still an outlier. As this case features high beam power and low beam torque, a possible explanation is that reduced $E\times B$ shearing leads to stronger turbulence levels and enhanced particle transport, which can advect particles and could produce an apparent momentum convection or generally overly flat density profiles. However, this remains a hypothesis, as no turbulence measurements are available for this discharge.

\begin{figure}
    \centering
    \begin{overpic}[width=0.4\linewidth]{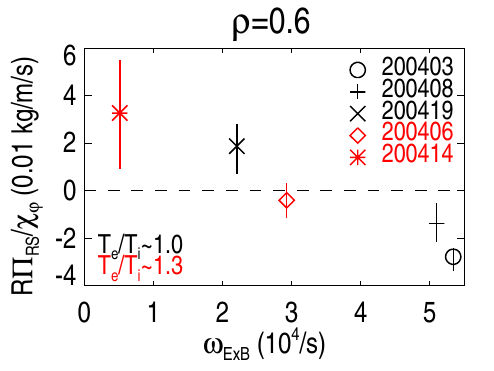}
    \put(55,20){\sffamily co-current}
    \put(40,55){\sffamily ctr.-current}
    \end{overpic}
    \caption{Normalized residual stress as a function of $E\times B$ shearing rate. Strong $E\times B$ shearing is associated with co-current intrinsic torque, while weak shearing tends to produce counter-current or near-zero torque. Shown is a reduced dataset with $T_e/T_i \approx 1.0-1.3$, selected to isolate the effect of modified background rotation while keeping the turbulence regime approximately constant.}
    \label{fig:ExB_scan}
\end{figure}

Next, Fig.~\ref{fig:TeTi_scan}(c) shows the normalized residual stress flux at $\rho_\varphi = 0.4$. Being strongly co-current directed in ITG conditions (low $T_e/T_i$), the residual stress contribution flattens toward values consistent with zero during the transition. For $T_e/T_i \gtrsim 1.2$, the data show a weak tendency toward co-current values again, which is within the experimental uncertainties of a flat trend. Panel (f) shows a clear non-monotonic, V-shaped trend at $\rho_\varphi = 0.6$: the residual stress is strongly co-current in the ITG regime, flattens or even reverses sign in the transition regime, and becomes co-current again in the deep TEM regime (high $T_e/T_i$). This behavior is reminiscent of trends previously observed on AUG \cite{Angioni_PRL_2011,Zimmermann_2024}. The observed trends can be correlated with kinetic gradients: Figure~\ref{fig:combined}(b) shows the residual stress number plotted over $R/L_{T_e}$, which is strongly cross-correlated with $R/L_{n_e}$ and orders most points well. This suggests that the observed V-shaped trend arises from profile shearing effects and the tilting of turbulent eddies as the turbulence regime shifts from ITG to TEM \cite{Camenen2011}. Earlier studies \cite{Angioni_PRL_2011, Grierson_PRL_2017, Hornsby2018, Zimmermann_2024} have shown that changes in the normalized density and electron temperature gradients, $R/L_{n_e}$ and $R/L_{T_e}$, play a major role in driving turbulent intrinsic torque via so-called profile shearing effects. From a theoretical perspective \cite{Camenen2011,Hornsby2018}, the generation of intrinsic torque should scale with the second derivatives of the kinetic profiles, which are in turn self-consistently set by the dominant turbulence regime \cite{Fable_2010, Angioni_2011_gyro,Fable_2019}. In practice, however, these higher-order derivatives are difficult to determine reliably from experimental measurements. As a result, analyses typically rely on the corresponding first-order gradients as practical proxies. A conclusive explanation of the experimental observations would ultimately require nonlinear, global gyrokinetic simulations \cite{Hornsby2018, Wang_PRL_2009, Wang_PoP_2010, Grierson_PRL_2017}. Performing such simulations lies beyond the scope of the present study.

In both Figs. \ref{fig:TeTi_scan}(f) and \ref{fig:combined}(b), low-rotation discharges \#200414 and \#200419 stand out, with strongly counter-current directed residual stress values at lowest $E \times B$ shearing. In an attempt to decouple different drivers of residual stress and understand the deviation of the low-rotation cases, Fig.~\ref{fig:ExB_scan} shows the normalized residual stress flux for datapoints with similar $T_e/T_i \approx 1.0$–$1.3$ plotted as a function of the $E \times B$ shearing rate. Both discharges designed to have low rotation values indeed feature the lowest shearing rates. This suggests that stronger $E\times B$ shearing results in co-current intrinsic torque generation, in agreement with theoretical expectations \cite{Dominguez_1993,garbet2002turbulence,Guercan_2007,Casson_PoP_2009,Staebler_PRL_2013}. This trend may be connected to previous intrinsic torque scalings at the edge of the plasma core, where the torque was found to scale with the ion pressure gradient as a proxy for $E \times B$ shearing. In the present work, however, the rotational contributions to the $E_r$ force balance are much larger than the differences in the pressure gradient, such that the $E \times B$ shearing emerges as a much more suitable ordering parameter. Overall, this explains the deviation of the low-rotation cases when plotted over $T_e/T_i$ or $R/L_{T_e}$. Differences in magnetic shear, which has also been suggested to affect the intrinsic torque \cite{Camenen2011}, are unlikely to explain the observed differences between the discharges, as the $q$ and shear profiles are very similar across the dataset.

\section{Comparison to Gyrokinetic Modeling}
\label{sec:GK_modeling}

\begin{figure}
    \centering
    \includegraphics[width=0.75\linewidth]{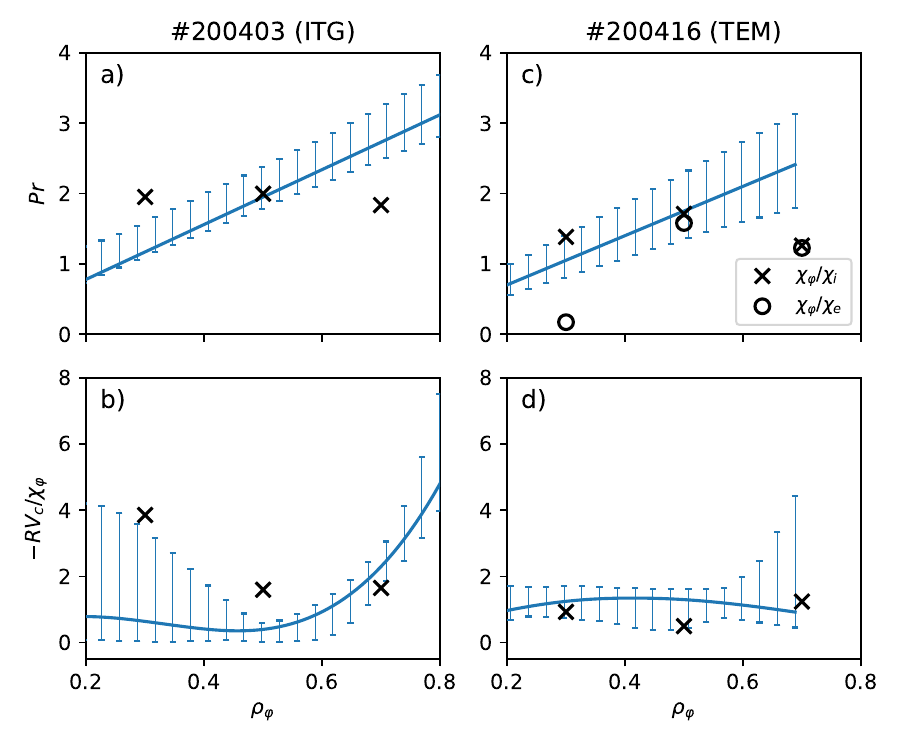}
    \caption{Comparison of CGYRO calculations (markers) against experimental analysis results (blue line with error bars). Shown are the ITG discharge \#200403 (left column) and the TEM discharge \#200416 (right column), with Prandtl numbers in the upper row and pinch numbers in the lower row. Overall, gyrokinetic predictions and experimental results are in good agreement for both discharges. The Prandtl numbers are slightly higher than previously observed on AUG for ITG-dominated plasmas. For the TEM discharge, Panel~(c) additionally shows the normalization of the momentum diffusivity to the electron heat diffusivity, $\chi_\varphi/\chi_e$, with the open circle marking the innermost radial point where $\chi_e$ significantly exceeds $\chi_i$, leading to a pronounced drop in $\chi_\varphi/\chi_e$.}
    \label{fig:GK_vs_exp}
\end{figure}

Beyond the detailed discussion of the observed experimental and theoretically expected parameter scans in the previous sections, the two most important discharges, the ITG discharge \#200403 and the deep TEM discharge \#200416, are studied more closely with the help of the CGYRO code. The goal is to compare linear gyrokinetic predictions for the pinch and Prandtl numbers with the experiment. Calculating the dominant contributions to the residual stress would require nonlinear calculations beyond the scope of this work.

Based on the CGYRO calculations discussed in Section \ref{sec:dataset}, simulations were run with experimental rotation and shearing rate, then repeated with either rotation or shearing rate artificially set to zero. Prandtl and pinch numbers were determined by comparing the momentum fluxes for these three rotation settings. A quasilinear estimate was used to obtain total fluxes at each radial location. Thereby, lowest $k_y \rho_s$ were found to contribute, in absolute units, most strongly to the overall spectral averaged fluxes. This workflow agrees with the one described in detail in Ref.~\cite{Zimmermann_dissertation} and as typically used for such predictions.

The results of these calculations are shown with markers in Fig.~\ref{fig:GK_vs_exp}, with the experimental fit shown in blue with error bars in the background. The Prandtl numbers are shown in the upper row and the pinch numbers in the lower row. For the ITG discharge \#200403, the overall average value for $Pr$ is in good agreement with the experimental estimate, see Panel~(a). Both the experiment and the prediction yield values higher (around $Pr \approx 2$) than previously predicted \cite{Zimmermann_2024}. This may result from the more consistent inclusion of carbon as a trace species in the CGYRO simulations performed here. Previous work only modified the collisional operator via $Z_\text{eff}$, which may not fully capture the stabilization of ion heat transport expected at modest carbon concentrations. This stabilization can occur primarily through dilution of the deuterium density and the consequent reduction in the free energy available to drive the mode \cite{Kotschenreuther_1995}. In this scenario, deuterium carries most of the ion heat flux while momentum transport is distributed more evenly between species, which could give rise to systematic differences in $Pr$ between the two modeling approaches. A more detailed comparison is left for future work. The corresponding estimates for the pinch number are shown in Panel~(b) and are mostly within the experimental error bars. While the innermost radial point of the prediction is somewhat elevated, the experimental error bars nearly encompass the corresponding predicted value. 

A similar picture holds for the TEM discharge \#200416. Panel~(c) shows overall good agreement between prediction and experiment for the Prandtl number. Again, the outermost radial point of the prediction is somewhat lower than the experimental estimate. This could indicate a larger ion heat diffusivity $\chi_i$ in the prediction, or an additional diffusive contribution in the experiment that is not captured by the gyrokinetic model. The pinch number predictions are in good agreement with the experimental values, as shown in Panel~(d).

Since in a deep TEM regime the electrons contribute more strongly to the energy fluxes, Panel~(c) also shows the normalization of the momentum diffusivity to the electron heat diffusivity, $\chi_\varphi/\chi_e$. For $\rho_\varphi = 0.5$ and $\rho_\varphi = 0.7$, this normalization yields very similar results to the typical normalization by $\chi_i$. Compared to previous theoretical work by Kluy~\textit{et al.}\ \cite{Kluy_2009}, the values obtained here are somewhat higher for $\chi_\varphi/\chi_e$. A possible reason is that $\chi_i$ and $\chi_e$ remain very similar in both the experiment and the gyrokinetic predictions across most radial locations, such that $\chi_i$ still remains a valid choice for normalization. This similarity does not hold for the innermost radial point at $\rho_\varphi=0.3$, where the predicted $\chi_e$ is substantially larger than $\chi_i$ and also larger than at the other radial locations, possibly being influenced by ETG turbulence at this radial position. In addition, the mixed-mode character of the turbulence and possible multi-scale interactions, which cannot be fully captured by the modeling approaches in this work, may also contribute. As a consequence, $\chi_\varphi/\chi_e$ drops significantly at this point, as indicated by the open circle marker, suggesting that a normalization with $\chi_e$ is likely not an appropriate measure in this case. Notably, this is also the data point with the lowest normalized density gradient $R/L_{n_e}$ in these simulations. In the work by Kluy~\textit{et al.}, the difference between $\chi_\varphi/\chi_i$ and $\chi_\varphi/\chi_e$ is largest at smaller values of $R/L_{n_e}$, which is consistent with the behavior observed here. In general, the kinetic profiles and their gradients are rather flat for \#200416 inside the ECH deposition (cp. Fig. \ref{fig:tglf_turbulence_regime}). Together with the very small heat fluxes, this makes the diffusivities a not very well defined quantity anymore.

Taken together with the results of the previous section, these findings support the elevated Prandtl numbers obtained in these specific cases, which exceed previously reported values. In general, the magnitude of both the Prandtl and pinch numbers is in good agreement between prediction and experiment for both the ITG and TEM discharges, suggesting the applicability of gyrokinetic predictions for assessing linear diffusive and convective momentum transport fluxes also in TEM-dominated regimes. As already discussed in the experimental analysis, uncertainties remain in the precise normalization chosen for the Prandtl number when sweeping across turbulence regimes. More broadly, the dataset from this experiment provides a wealth of opportunities for future gyrokinetic studies. These include investigations of the influence and the implementation of the carbon impurity background, a closer study of cases with lower experimental $Pr$ values, and potentially the residual stress itself through nonlinear gyrokinetics.

\section{Discussion and Summary}
\label{sec:summary}

This work extends the momentum transport analysis framework established on AUG to \diiid{} plasmas. The analyzed NBI modulation experiments cover a transition from ITG-dominated to deeply TEM-dominated turbulence, while also varying the input torque and, consequently, $E\times B$ shearing. The dataset therefore approaches reactor-relevant conditions more closely than previous studies, combining dominant electron heating with low external torque and, in some cases, $E\times B$ shearing rates comparable to the turbulence growth rates.

A first important result is methodological: the analysis framework previously developed for AUG could be successfully transferred to \diiid{} and reproduced the measured steady-state, amplitude, and phase profiles reasonably well across the dataset. Also notable is the finding that results with and without co-/counter-NBI fall generally onto the same trend lines.

The Prandtl numbers are found to show an unexpected variation outside mid-radius, suggesting that additional effects not included in the present model may affect its fitting. In particular, the common assumption that momentum diffusivity scales primarily with ion heat diffusivity may become insufficient when the turbulence is increasingly electron-driven. In such a scenario, this method may be approaching its limits as electron turbulence increases, rotation perturbations become relatively smaller, and heating perturbations possibly become larger. Future modeling work should perform a more detailed analysis of the cases with lower Prandtl numbers. Notably, some of the Prandtl numbers obtained on \diiid{} are higher than previously predicted. However, the linear gyrokinetic predictions for both the Prandtl and pinch numbers are in good agreement with the experimentally inferred values for the modeled cases, which belong to the higher-$Pr$ group. This may be related to the inclusion of a carbon impurity background as its own species in the simulations, representing a higher degree of realism compared to previous work. Future work should perform a more detailed comparison of this effect in the modeling. In general, these results suggest that linear gyrokinetic modeling is applicable for assessing diffusive and convective momentum transport also in TEM-dominated regimes, while the precise role of the Prandtl number normalization across turbulence regime transitions remains an open question for future work. 

In the experiment, the normalized pinch number does not show an explicit dependence on the ITG-TEM transition. Instead, it is roughly ordered by the self-consistently emerging density gradient, consistent with expectations from the Coriolis pinch, while the low-rotation, weak $E\times B$ shearing cases appear as outliers with enhanced apparent convection. 

A clear trend is found for the normalized residual stress. It exhibits a non-monotonic, V-shaped dependence when ordered by $T_e/T_i$: co-current in deep ITG, reduced or counter-current in the intermediate mixed-mode regime, and co-current again in the deep TEM cases. This behavior is consistent with previous observations on AUG and suggests that the transition region between ITG and TEM is the regime in which intrinsic torque is most likely to flatten or reverse, whereas fully developed TEM turbulence can again support co-current intrinsic torque. A key result is that this V-shaped trend collapses onto an approximately linear dependence when plotted against $R/L_{T_e}$. This strongly suggests that the residual stress is not controlled by $T_e/T_i$ or the turbulence regime itself, but rather by the kinetic-profile gradients that evolve together with the turbulence regime. The results are therefore consistent with profile-shearing-based symmetry breaking as the main mechanism setting the sign and magnitude of the turbulent intrinsic torque. In addition, the background $E\times B$ shearing shifts the residual stress level. At comparable $T_e/T_i$, stronger $E\times B$ shear is associated with more co-current residual stress, while weak-shear cases move toward counter-current values. This is particularly relevant for reactor extrapolation, since future devices are expected to operate with low external torque and therefore weaker rotational shear and intrinsic torque drive.

In conclusion, for a future reactor scenario in which rotational shearing and peaked core rotation profiles are desirable for stable operation, a mixed-mode turbulence regime appears to be least favorable for building up strong core rotation gradients. In contrast, both deep ITG and deep TEM regimes are associated with more co-current residual stress contributions, which can support the formation of peaked rotation profiles. At the same time, peaked pressure profiles may enhance $E \times B$ shearing through the radial electric field, which could further contribute to rotational peaking and partially compensate for weaker intrinsic torque in mixed-turbulence regimes. 

\appendix

\section{Main Ion Measurements}
\label{sec:appendix_main}

\begin{figure}
    \centering
    \includegraphics[width=0.4\linewidth]{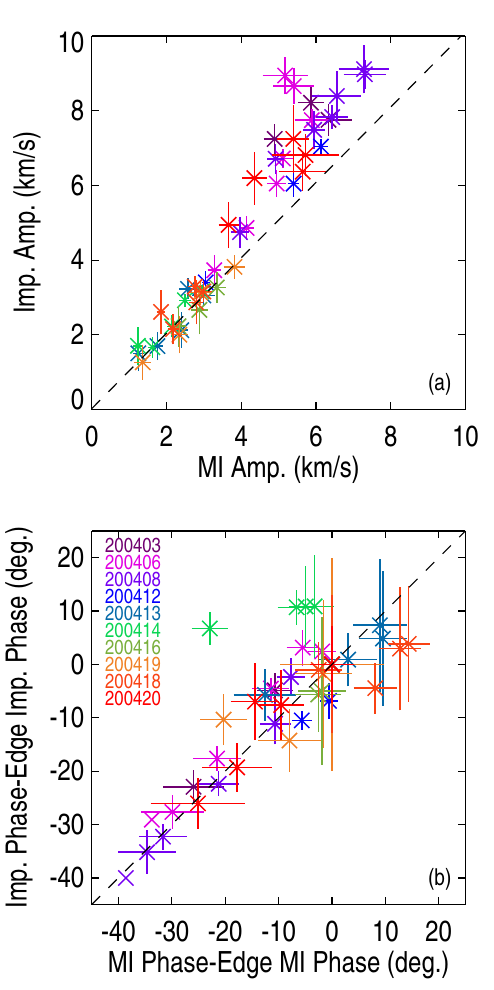}
    \caption{Comparison of the rotation modulation (a) amplitude and (b) phase for main-ion and impurity views in the key discharges for this experiment. Each point represents a comparison between an impurity and main-ion view with very similar measurement locations. The phase measurements are made relative to the impurity and main-ion measurement made closest to the plasma edge.}
    \label{fig:imp_vs_main_ion}
\end{figure}

As noted earlier, all the measurements of toroidal rotation used in this work come from the carbon impurity in the plasma, while the bulk of the momentum is carried by the deuterium ions in the plasma (the main ions). Many studies of momentum transport, including this one, rely on the assumption that collisional coupling between the impurities and main ions makes their rotations nearly identical, and past experiments tend to only observe small deviations from this, especially in the core of the plasma \cite{grierson_2012,Grierson_2013}. However, an experiment like this one, which relies on precise time dynamics of rotation across conditions that vary the collisionality, benefits from additional scrutiny of this assumption. Although \diiid{} has a mature Main Ion CER (MICER) diagnostic system, the experiment's need for continuous rotation measurements prevents the most precise MICER analysis from being done. This is because the passive component of the MICER spectrum is very dynamic and essential to remove in order to do precise analysis (much more so than for the impurity CER analysis), and this requires modulating the neutral beam injection in a way that prevents the continuous measurement of impurity rotation. Despite this limiting the reconstruction of the absolute profile height, analysis with sufficient accuracy to obtain the amplitude and phase of the modulated main-ion rotation was possible for the key discharges in this experiment. The comparison of the impurity and main-ion modulation amplitude (for several views across the core of the plasma) is shown in Fig.~\ref{fig:imp_vs_main_ion}(a) and Fig.~\ref{fig:imp_vs_main_ion}(b) shows the comparison of the phase of the rotation modulations (measured relative to the outermost impurity and main-ion measurement, respectively). Here, it can be seen that the agreement is generally good. The impurity amplitudes are slightly higher than the main-ion amplitudes (for the higher amplitude cases), but the difficulty of the MICER analysis in this case is a potential root cause of this difference. The phase results are all in quite good agreement, aside from a few points for 200414; however, this discharge has some of the lowest amplitude modulations (reflected in the large uncertainties in the fitting solution above), and, again, the difficulty of the main-ion analysis precludes drawing any strong conclusion based on this result. Considering these results in concert with an expectation that the mean toroidal rotation will be very similar for the impurities and the main ions (based on past MICER results, see, for example Ref. \cite{Chrystal2017}), it is expected that the momentum transport results presented in this work are largely representative of the bulk plasma momentum transport.

\section{ECH Power Deposition per Gyrotron}
\label{app:ech}

\begin{figure}
    \centering
    \includegraphics[width=0.7\linewidth]{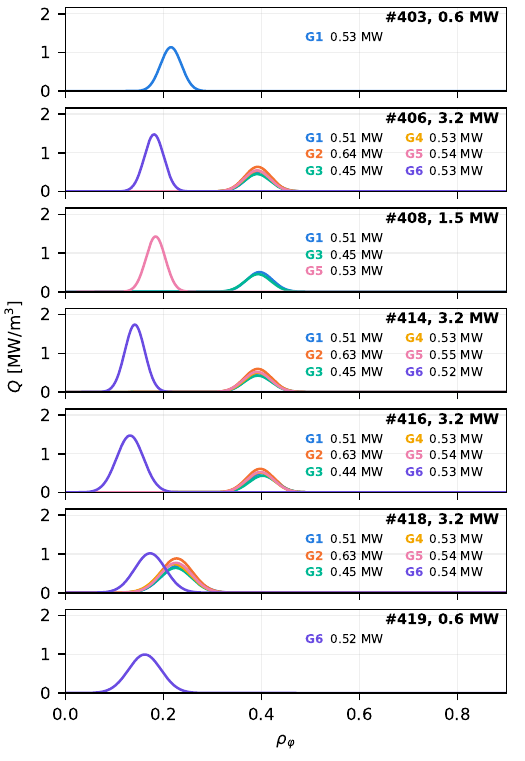}
    \caption{ECH power deposition profiles of the individual gyrotrons for all discharges as calculated via TORBEAM in TRANSP. The numbers in each panel list the absorbed power of every gyrotron in MW.}
    \label{fig:ech_gyrotrons}
\end{figure}

ECH is the main actuator used in these experiments to steer the dominant turbulence regime, and its deposition location was varied between the discharges depending on experimental and operational requirements, in particular to avoid strong core MHD activity. Fig.~\ref{fig:ech_gyrotrons} shows the ECH power deposition of the individual gyrotrons for all discharges listed in Table~\ref{tab:shotlist}. The profiles are taken from the TORBEAM runs in TRANSP, averaged over the respective analysis windows. For all cases, the ECH power is concentrated in the inner half of the plasma, and inside or at the radii where the primary results are reported.

\section*{Acknowledgments}
The authors would like to acknowledge the support and helpful discussions with Kathreen Thome, Alessandro Bortolon, and Christopher Holland. Part of the data analysis was performed using the OMFIT integrated modeling framework \cite{Meneghini_2015}. This work has been carried out within the framework of the EUROfusion Consortium, funded by the European Union via the Euratom Research and Training Programme (Grant Agreement No 101052200 — EUROfusion). Views and opinions expressed are however those from the author(s) only and do not necessarily reflect those of the European Union or the European Commission. Neither the European Union nor the European Commission can be held responsible for them. This material is based upon work supported by the U.S. Department of Energy, Office of Science, Office of Fusion Energy Sciences, using the DIII-D National Fusion Facility, a DOE Office of Science user facility, under Award(s) DE-FC02-04ER54698, DE-AC02-09CH11466, DE-FG02-08ER54999, DE-SC0020287, DE-SC0026404, DE-SC0022270 and DE-GG020277. This report was prepared as an account of work sponsored by an agency of the United States Government. Neither the United States Government nor any agency thereof, nor any of their employees, makes any warranty, express or implied, or assumes any legal liability or responsibility for the accuracy, completeness, or usefulness of any information, apparatus, product, or process disclosed, or represents that its use would not infringe privately owned rights. Reference herein to any specific commercial product, process, or service by trade name, trademark, manufacturer, or otherwise does not necessarily constitute or imply its endorsement, recommendation, or favoring by the United States Government or any agency thereof. The views and opinions of authors expressed herein do not necessarily state or reflect those of the United States Government or any agency thereof. The Swiss contribution to this work has been funded by the Swiss State Secretariat for Education, Research and Innovation (SERI). Large Language Models were used during the preparation of this manuscript to assist with grammar checking and to improve readability. All scientific content, interpretations, and conclusions have been carefully verified by the authors. 

\printbibliography

\end{document}